\documentclass[twocolumn, superscriptaddress, amsmath, amssymb, floatfix, notitlepage, nofootinbib, 10pt, reprint, prl]{revtex4-1}
\usepackage[latin9]{inputenc}
\usepackage{array}
\usepackage{amsmath}
\usepackage{bm}
\usepackage{braket}
\usepackage{booktabs}
\usepackage{color}
\usepackage{dsfont}
\usepackage{dcolumn}
\usepackage{esint}
\usepackage{epstopdf}
\usepackage{graphicx}
\usepackage{mathrsfs}
\usepackage{tabu}
\usepackage{graphicx}
\usepackage{enumitem}

\usepackage[colorlinks=true,citecolor=blue,linkcolor=blue]{hyperref}

\newcommand{\bea}{\begin{eqnarray}}
\newcommand{\eea}{\end{eqnarray}}
\newcommand{\bt}{\textbf}

\newcommand{\noi}{\noindent}
\newcommand{\no}{\nonumber}

\newcommand{\appropto}{\mathrel{\vcenter{
  \offinterlineskip\halign{\hfil$##$\cr
    \propto\cr\noalign{\kern2pt}\sim\cr\noalign{\kern-2pt}}}}}
\begin{document}

\title{Field-Programmable Topological Array: Framework and Case-Studies}

\author{G.-Y.~Huang}
\affiliation{Institute for Quantum Information \& State Key Laboratory of High Performance Computing, College of Computer Science, NUDT, Changsha 410073, China}

\author{B.~Li}
\affiliation{Institute for Quantum Information \& State Key Laboratory of High Performance Computing, College of Computer Science, NUDT, Changsha 410073, China}

\author{X.-F.~Yi}
\affiliation{Institute for Quantum Information \& State Key Laboratory of High Performance Computing, College of Computer Science, NUDT, Changsha 410073, China}

\author{J.-B.~Fu}
\affiliation{Institute for Quantum Information \& State Key Laboratory of High Performance Computing, College of Computer Science, NUDT, Changsha 410073, China}

\author{X.~Fu}
\affiliation{Institute for Quantum Information \& State Key Laboratory of High Performance Computing, College of Computer Science, NUDT, Changsha 410073, China}

\author{X.-G.~Qiang}
\affiliation{Institute for Quantum Information \& State Key Laboratory of High Performance Computing, College of Computer Science, NUDT, Changsha 410073, China}
\affiliation{National Innovation Institute of Defense Technology, AMS, Beijing 100071, China;}

\author{P.~Xu}
\affiliation{Institute for Quantum Information \& State Key Laboratory of High Performance Computing, College of Computer Science, NUDT, Changsha 410073, China}

\author{J.-J.~Wu}
\affiliation{Institute for Quantum Information \& State Key Laboratory of High Performance Computing, College of Computer Science, NUDT, Changsha 410073, China}

\author{C.-L.~Yu}
\affiliation{China Greatwall Quantum Laboratory, Changsha 410006, China}

\author{P.~Kotetes}
\email[Corresponding author: ]{kotetes@itp.ac.cn}
\affiliation{CAS Key Laboratory of Theoretical Physics, Institute of Theoretical Physics, Chinese Academy of Sciences, Beijing 100190, China}

\author{M.-T.~Deng}
\email[Corresponding author:]{mtdeng@nudt.edu.cn}
\affiliation{Institute for Quantum Information \& State Key Laboratory of High Performance Computing, College of Computer Science, NUDT, Changsha 410073, China}

\begin{abstract}
Engineering composite materials and devices with desired topological properties is accelerating the development of topological physics and its applications. Approaches of realizing novel topological hybrids, including \textit{in-situ} epitaxy growth, planar/layered superlattices, and assembled (artificial) atom/dot arrays, etc., endow the topological systems with a substantial degree of control and tuna\-bi\-li\-ty. Here, we propose a framework for realizing a field-programmable topological array (FPTA) that enables the implementation of dynamically reconfigurable topological platforms. FPTA allows for the independent, simultaneous, and local programmability of the various platform properties, such as the electromagnetic field, the spin-orbit field, and the superconducting order parameter. To demonstrate the effectiveness of the FPTA in rendering the system topologically-nontrivial and implementing non-Abelian manipulations, we simulate their operation in various case-studies. Our framework provides a playground for unearthing novel topological phases using components of high feasibility and sets the guidelines for run-time dynamic reconfiguration which is crucial for high-performance topological electronic circuits and quantum computing.
\end{abstract}

\date{\today}

\maketitle

%============================###########===========================
%============================###########===========================

\section{Introduction}\label{sec:SectionI}

Topological order is one of the most intensely investigated topics lying on the frontiers of condensed matter physics and materials science~\cite{Hasan2010, Qi2011}. These phases go beyond Landau's classification of matter, which relies on a set of order parameters that characterize the spontaneous symmetry breaking taking place~\cite{Landau1980}. Their unconventional behavior spurred the development of a broader symmetry classification of phases of matter~\cite{Altland,Schnyder,KitaevClassi,Ryu}, where the so-called topological invariants~\cite{Thouless1982} are employed instead of the Landau order parameters. Topological band insulators and semimetals possess unique electronic band structures, which lead to protected boundary states that appear critical for advancing high-performance electronic devices~\cite{Liu2016} and robust quantum computing systems~\cite{Nayak2008,Pachos2012}. 

For most materials, however, an intrinsic topological state is either inaccessible or very poorly exploitable for applications. Therefore, nontrivial topology is often pursued by means of artificial engineering. The application of ultra-clean \textit{in-situ} epitaxial growth technologies, which provide an atom-by-atom crystal assembly, has greatly boosted the materialization of various to\-po\-lo\-gi\-cal orders~\cite{Ginley2016}. Among them, one finds the representative examples of quantum spin~\cite{Konig2007} and anomalous~\cite{Chang2013} Hall insulators, Weyl semimetals~\cite{Xu2015,Lv2015}, and Majorana zero mode (MZM) platforms~\cite{Mourik2012, Perge2014, Yin2015, Deng2016, Wang2018, Liu2018}. By following the alchemistic databases of topological quantum chemistry~\cite{Zhang-Vergniory-Tang2019} and loading the furnaces with the right chemical elements, a plethora of exotic topological phases of matter appear to be on the brink of discovery. 

One of the ultimate goals of this field is the design and experimental demonstration of manipulable to\-po\-lo\-gi\-cal devices. However, this pursuit is still facing many challenges that need to be conquered. The principal hurdle in this effort is the difficulty in subjecting topological materials to nanofabrication processes without inducing fatal defects. The majority of the currently accessible topological materials are sensitive to ambient conditions (oxygen, moisture and pressure, etc.), which are factors that are hard to control in realistic device fabrications. Further, some topological material families are naturally difficult to tune using traditional electrical gates, e.g., metallic superconductor based systems~\cite{Perge2014,Wang2018, Liu2018}. 

%%%%%%%%%%%%%%%%%%%%%%%%%%%%%%%%%%%%%%%%%%%%%%%%%%%%%
%---------------------------------------------------------------------------------------------------------------------------------
\begin{figure*}[ht]
\centering \includegraphics[width=17 cm]{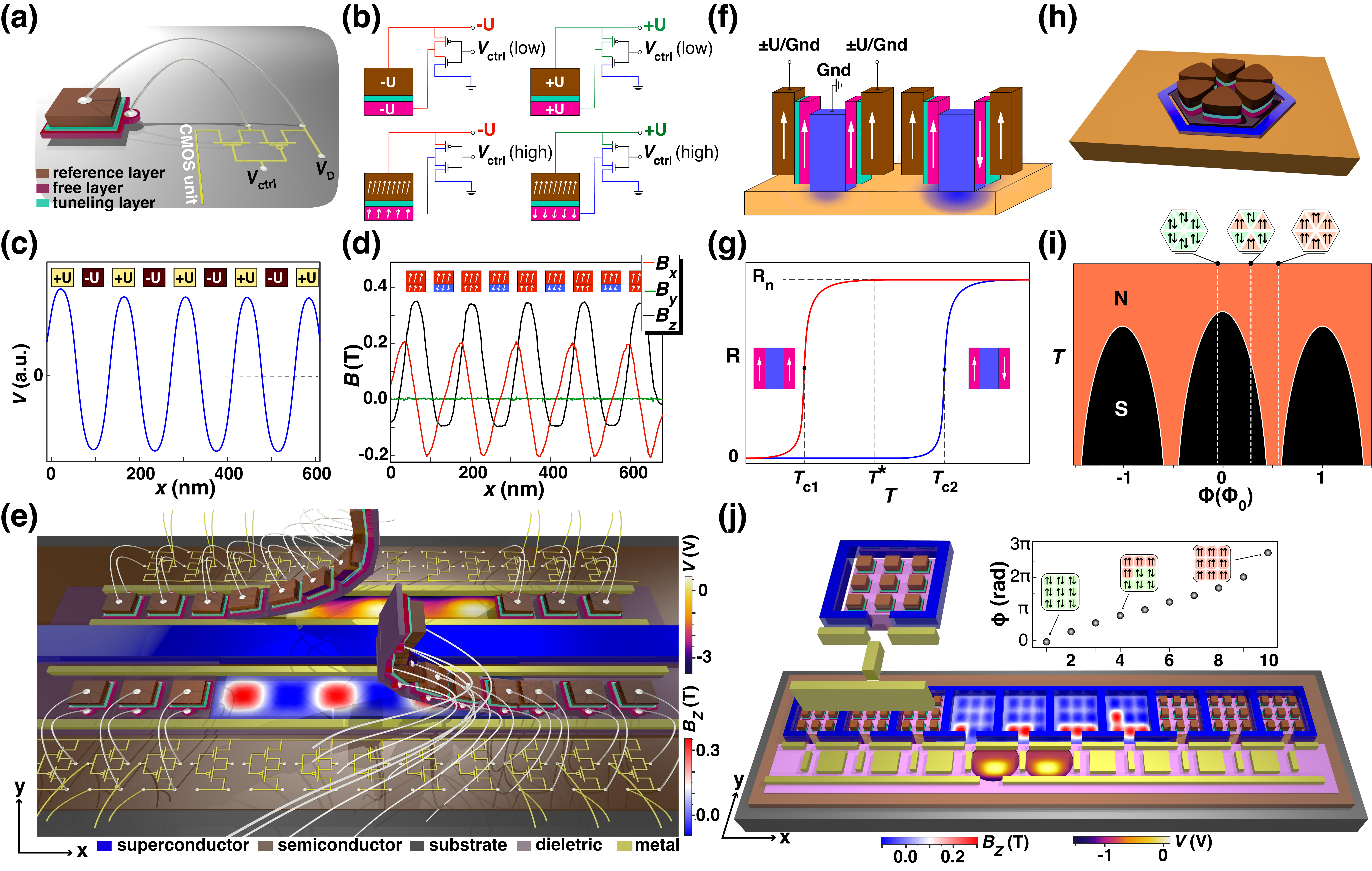}
\caption{\label{Fig1} Design of the field-programmable topological array (FPTA) control cells. (a) Conceptional schematic of an FPTA cell for local gate and magnetic field controls. Here, the cell is mainly composed of an STT-cell structure and a controlling CMOS unit. Note that the STT structure illustrated in this paper is simplified for convenience. (b) The FPTA cell can be configured into voltage or magnetic control modes by the auxiliary CMOS unit. (c) and (d) Effective voltage and magnetic field underneath an FPTA array, respectively. The calculation is performed using the FEA software COMSOL Multiphysics~\cite{comsol}. (e) A double-array configuration involving the FPTA control cells in panels (a) and (b). (f) an FPTA cell design composed of two STTs and one SSV, which implements a superconductivity switch. (g) Operating principle of the superconductivity switch in (f). Depending on whether the sandwiching ferromagnet layers are polarized in a parallel or an anti-parallel fashion, the critical temperature of the superconductor can be significantly modified. (h) Design for a superconducting vorticity FPTA control cell, which is assembled from STT clusters and a doubly connected superconductor. (i) Operating principle of the vorticity controller. The stray-field induced flux through the superconductor can change the phase winding number and therefore the vorticity. (j) A single superconducting quantum-dot-chain configuration. The STT clusters are used to tune the phase difference between the adjacent superconductor dots. The inset plot of (j) shows the simulated flux acquired from the $3\times3$ STT matrices.}
\end{figure*}
%---------------------------------------------------------------------------------------------------------------------------------
%%%%%%%%%%%%%%%%%%%%%%%%%%%%%%%%%%%%%%%%%%%%%%%%%%%%%

In order to overcome these challenging obstacles, va\-rious methods have been proposed and developed, which designate how to obtain topological heterostructures by gluing together well-studied materials. The resulting composite system can host easy-to-manipulate to\-po\-lo\-gi\-cal phases, which are not intrinsically supported by the parent materials. For instance, topological insulators, or semiconductors with large spin-orbit interaction (SOI), are predicted to harbor manipulable MZMs~\cite{Fu2008,Sau2010} when they are in proximity to s-wave superconductors. Even more, one can also synthesize new topological hybrids by stacking different materials into superlattice structures~\cite{Burkov2011}, which become coupled either via chemical~\cite{Shibayev2019} or van-der-Waals~\cite{Geim2013,Cao2018} bonds. Nowadays, nanofabrication techniques even enable us to define planar superlattices directly. This opens perspectives for emulating various many-body phenomena~\cite{Singha2011}, as well as give access to topological phases in more flexible and customized artificial crystalline structures~\cite{Sushkov}.

Here, motivated by the abovementioned landmark works on topological artificial crystals, we take a na\-tu\-ral leap forward and propose a framework for a field-programmable topological system. This approach relies on the deposition of an array of control cells on top of a target material substrate. The resulting hybrid system allows the real-time induction of multiple topological phases in different regions, which are spatially defined by solely controlling a set of electrostatic gates. This system is here-coined \textit{field-programmable topological array} (FPTA). The present manuscript does not aim at pro\-vi\-ding a complete and comprehensive theory for the FPTA. Instead, it targets to widely expose this promising expe\-ri\-men\-tal approach and sti\-mu\-late the research on digitized topological systems.

The remainder of this manuscript is organized as follows. In Sec.~\ref{sec:SectionII}, we discuss the design of the FPTA control cells. In Sec.~\ref{sec:SectionIII}, we explore four case-studies (\bt{A}-\bt{D}), where FPTAs are used to control electron transport phenomena (\bt{A}), ena\-ble topological phase transitions (\bt{B}), implement non-Abelian braiding (\bt{C}), and engineer synthetic Weyl points (\bt{D}). Finally, we discuss our conclusions in Sec.~\ref{sec:SectionIV}.

\section{Design of FPTA control cells}\label{sec:SectionII}

In order to engineer the desired topological phase and manipulate its properties in a spatiotemporally controllable manner, it is essential to employ microsized control units with high (re-)configurability. The state-of-the-art lithography can provide high density and uniform multi-function control matrix assembled by stable transistor units, spintronic units, ferroelectric units, or superconducting units.

The most straightforward control-cell structure, especially for semiconductor-based platforms, is a gate electrode. It is well-established that a global gate can tune the position of the Fermi level and thus control the topological phases~\cite{Deng2020, Liu2020}. Recently, it was pointed out that laterally-patterned local electrode arrays~\cite{Sushkov} could effect topological-phase transitions by tuning the electron density distribution or other material parameters (e.g. Rashba SOI~\cite{Wojcik2018, Liang2012}). 

Apart from electric fields, magnetic fields also allow controlling topological phase transitions. However, this is possible by means of violating time-reversal symmetry which is effected either through flux threading or Zeeman spin splitting. To achieve versatile magnetic-field tuning one can employ micro-magnets, which allow for spatially-resolved magnetic field control. Con\-si\-de\-ring independently-configurable micro-magnets instead of a uniform magnetic field, promises an enhanced degree of tunability. For example, micro-magnets can produce a magnetic field with an oscillating amplitude, which is able to generate a synthetic SOI~\cite{BrauneckerSOC} and remedy the lack of it in some topological phase realizations~\cite{Kjaergaard2012, Zutic, Kontos}.

To showcase the possibility of the combined control of electric and magnetic fields in the FPTA approach, we depict in Figs.~\ref{Fig1}(a) and~(b) the design of an electromagnetic control cell. In this case, both fields can be controlled in a local fashion by utilizing a spin-transfer-torque (STT) device~\cite{Slonczewski1997, Berger1996}, or alternatively, a spin-orbit-torque (SOT) device~\cite{Chernyshov2009}. As depicted in Figs.~\ref{Fig1}(a) and~(b), when the control voltage applied on an auxiliary CMOS unit, i.e. $V_{\rm ctrl}$, is configured to a ``low" value, the free and reference layers of the STT-cell are both connected to the external voltage $U$, and the STT block serves as a local gate. Instead, when $V_{\rm ctrl}$ is switched to a ``high" value, the STT free layer is grounded. Depending on whether the external voltage $U$ is negatively or positively biased, the induced current between the free layer and the reference layer can set the STT to either ``parallel" or ``anti\-pa\-ral\-lel" spin polarization. The stray-field of the STT-cell is therefore configurable, as this is also verified by the simulated local voltage and stray-field shown in Figs.~\ref{Fig1}(c) and~(d). 

Featuring an access time of the nanosecond order, and over a 10\textsuperscript{12}-cycle operation endurance, the STT-cell can be tuned to various configurations with a high-level of programmability. For instance, Fig.~\ref{Fig1}(e) demonstrates a double-1D array configuration based on a semiconductor-2DEG/superconductor hybrid substrate~\cite{Shabani2016}. The STT-cells are arranged on top of the exposed semiconductor-2DEG and control the local electric and magnetic field profiles independently. See the exposed textures of the simulated electric and magnetic field color scales. In Sec.~\ref{sec:SectionIII}, we elaborate on concrete examples where topological phase transitions are effected by means of STT-cells. We remark that recent works in Ref.~\onlinecite{Zutic} have also considered STT-cells for topological-phase engi\-nee\-ring. However, the FPTA methodology presented here goes beyond these works, since the STT-platform combine electric field, Zeeman field, SOI-field and magnet flux tunabilities in this paper. Moreover, the STT is only one of the possible control units that can be implemented within this framework. 
 
In addition, within the present approach, we also consider previously~\cite{Zutic} not examined functionalities of the basic STT-cells, such as using them to control the superconducting order parameter. There are several examples of using an electric gate to tune the strength of the supercon\-duc\-ting proximity effect in quantum devices~\cite{Su2017, Rasmussen2018}. Even more, the STT-cells even allow for the direct tuning of the superfluid density, for instance, via superconducting spin valve (SSV) structures~\cite{Oh1997, Tagirov1999}. As shown in Figs.~\ref{Fig1}(f) and~(g), a superconductor sandwiched between two STT (or SOT) layers prepared in the ferromagnetic regime, can toggle between a higher- and a lower-T$_c$ (or even a normal state). 

Another route to manipulate superconductivity is to take advantage of the Little-Parks effect~\cite{LittleParks1962}. In Figs.~\ref{Fig1}(h) and~(i), we show a Little-Parks control-cell which is composed of an STT-matrix and an encircling superconducting ring. The stray-field of the STT-matrix induces an enclosed flux $\Phi$, whose amplitude depends on the configuration of the STT-matrix. To preserve the quantization of the fluxoid, the Cooper-pair condensate circulates in the ring, thus producing a net current that modifies the critical temperature. If the ring size is comparable to the superconducting coherence length, the ring undergoes a phase transition and enters into a complete destructive regime near a half-integer-flux values~\cite{Liu2001, Vaitiekenas2020}. Moreover, the Little-Parks cell is also able to directly tune the vorticity, therefore allowing the programming of vortex lattice systems~\cite{Biswas2013} or implementing braiding operations in vortex-MZM platforms~\cite{Wang2018, Liu2018}. In a si\-mi\-lar fashion, we can controllably adjust the phase dif\-fe\-ren\-ces in an array of local superconductors by employing the same approach as in the Little-Parks cell as shown in Fig.~\ref{Fig1}(j). This method can be harnessed to manipulate magnetic-field-induced topological phases~\cite{Fulga2013, Fornieri2019, Heck2014, Yang2019}.

We could enumerate more examples of FPTA-cell designs beyond the abovementioned spintronics and superconducting elements. For instance, magnonic~\cite{Krawczyk2014}, piezoelectric~\cite{McKinstry2018}, or other micro-structures are eligible, as long as these can be accessed fast, independently, repeatedly, and can provide a sufficiently high degree of control over the target material property of interest. 

%============================###########===========================
%============================###########===========================

\section{FPTA Case-Studies}\label{sec:SectionIII}

In this section, we show four case-studies using STT-based FPTA to control basic electron transport, topological phase transition, and non-Abelian operations in semiconductor-superconductor (SM-SC) hybrid system.

%============================###########===========================
%============================###########===========================

\subsection{FPTA-Mediated Crossed-Andreev-Reflection Enhancement}\label{sec:SectionIII_A}

In the first case-study, we demonstrate the capabilities of the FPTA framework, through the study of the crossed-Andreev-reflection (CAR) enhancement. 

CAR is a scattering process which takes place when at least two normal leads are simultaneously attached to a superconductor. CAR effects the transformation of an incoming particle (hole) of a given lead, into an out\-going hole (particle) in another lead, via the absorption (emission) of a Cooper pair. CAR should be contrasted to the local-Andreev reflection (LAR) process, where the particle-hole conversion occurs via a retro-reflection event in the same lead. This remarkable difference allows us to employ CAR for splitting a spin-singlet Cooper pair into a pair of highly-entangled and spatially-separated electrons. Achieving such entangled non-local states is fundamentally crucial for quantum information applications~\cite{Loss98}. However, CAR is hard to observe in a conventional metal/superconductor/metal setup, due to its weak strength. Many proposals providing distinct routes to enhance the CAR amplitudes have been already put forward. A number of them rely on ferromagnetic leads~\cite{apl76.487, prl93.197003}, graphene/silicene/MoS$_{\rm 2}$ devices~\cite{prl100.147001, prb85.035402, prl103.167003, prb94.165441, prb90.195445}, or topological materials~\cite{prl101.120403, prl109.036802, prl110.226802, prl122.257701}. 

\begin{figure}[t]
\centering \includegraphics[width=8.5 cm]{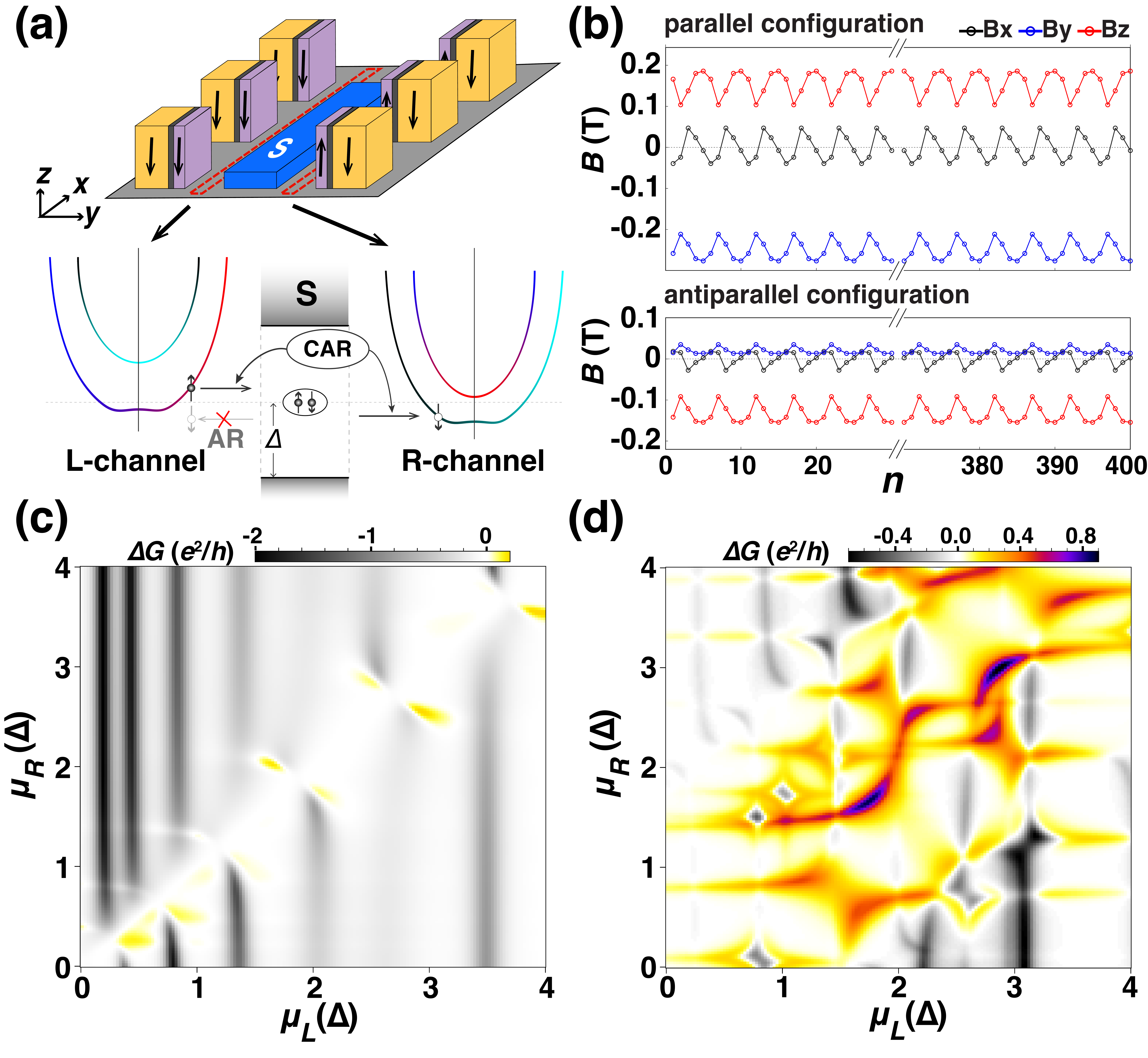}
\caption{\label{Fig2} FPTA approach for enhancing the crossed Andreev reflection (CAR). (a) The schematic of a related transport setup involving an FPTA and the corresponding CAR mechanism proposed here. (b) The magnetic field profiles of STT-cells with the lower (upper) panel possessing an (anti)parallel configuration. (c)-(d) The difference of the li\-near conductance $\Delta G\equiv G_{\rm CAR}-G_{\rm LAR}$ as a function of the chemical potentials of the two 1D channels $\mu_{\rm L}$ and $\mu_{\rm R}$ without STT-cells [(c)] and with STT-cells [(d)]. The parameter values used in the tight-binding calculations read: $t=20$, $\mu_{\rm sc}=30$, $t_{\rm L}=t_{\rm R}$, $\Gamma_{\rm L}=\Gamma_{\rm R}=0.1$, all in units of $\Delta$. The area of the central region consists of $60\times5$ sites, and the lattice spacing is $a\approx 10{\rm nm}$.} 
\end{figure}

Here, we demonstrate that the amplitude for the CAR can be greatly enhanced in a conventional normal/superconductor/normal system by utilizing an STT-based FPTA. The studied system is shown in the upper panel of Fig.~\ref{Fig2}(a), where a quasi-1D superconductor is sandwiched by two gate-defined 1D-electron gas channels (L-channel \& R-channel). Each channel is within the reach of the stray field induced by the nearby STT-array. To increase the amplitude for CARs, the STT-array near the L-channel is configured in a parallel fashion, while the STT-array near the R-channel sees an anti-parallel configuration. For a sketch, see the lower panel of Fig.~\ref{Fig2}(a). In such a setup, the electron spins near the Fermi level of the two channels are polarized in an antiparallel manner. It is shown that it can increase the CAR amplitude greatly. In contrast, the process of LAR is mostly prohibited due to the opposite spin-polarization on each side.

%%%%%%%%%%%%%%%%%%%%%%%%%%%%%%%%%%%%%%%%%%%%%%%%%%%%%
%---------------------------------------------------------------------------------------------------------------------------------
%--------------------------------------------------------------------------------------------------------------------------------
%%%%%%%%%%%%%%%%%%%%%%%%%%%%%%%%%%%%%%%%%%%%%%%%%%%%%

In order to quantitatively describe the FPTA-induced enhancement of the CAR amplitude, we numerically investigate a tight-binding model in conjunction with a finite-element analysis (FEA). The total Hamiltonian of the system consists of three parts, i.e., $H_{\rm total}=H_{\rm ch}+H_{\rm sc}+H_{\rm c}$, where $H_{\rm ch}$ is the Hamiltonian of the 1D channels, $H_{\rm sc}$ is the Hamiltonian of the superconductor, and $H_{\rm c}$ is the coupling Hamiltonian between the electronic channels and the superconductor. $H_{\rm ch}$ reads:
\bea
H_{\rm ch}=
\frac{1}{2}\sum_{n,\eta}
\left[c_{n,\eta}^\dag{\cal H}_\eta^Ac_{n,\eta}+
\Big(c_{n,\eta}^\dag{\cal H}_\eta^Bc_{n+1,\eta}+
{\rm H.c.}\Big)\right],\quad
\label{eq:Hch}
\eea

\noi where $c_{n,\eta={\rm L,R}}^{\dag}=(c_{n\uparrow,\eta}^{\dag},c_{n\downarrow,\eta}^{\dag},c_{n\downarrow,\eta},-c_{n\uparrow,\eta})$ defines the Bogo\-liu\-bov de Gennes (BdG) spinor, with $c_{n\sigma,\eta}^\dag$ ($c_{n\sigma,\eta}$) the electron creation (annihilation) operator in the $\eta$-channel at site $n$ with spin projection $\sigma$. The matrices ${\cal H}_\eta^{A,B}$ are given as:
\bea
{\cal H}_\eta^A=\big(2t-\mu_\eta\big)\tau_z
+\bm{{\cal V}}_{n,\eta}\cdot\bm{\sigma}\quad{\rm and}\quad
{\cal H}_\eta^B=-t\tau_z,\quad
\eea

\noi in which, $t={\hbar^{2}}/{2m_e^*a^2}$ is the hopping amplitude, with $\hbar$ the reduced Planck constant, $m_e^*$ the effective electron mass in the channels, $a$ the lattice spacing, and $\mu_\eta$ the chemical potential of the $\eta$-channel. In the above, $\boldsymbol{\mathcal V}_{n,\eta}\cdot\bm{\sigma}$ denotes the local Zeeman contribution to the energy felt by the $\eta$-channel at site $n$, where $\boldsymbol{\mathcal{V}}_{n,\eta}=\frac{1}{2}g\mu_B\bm{B}_{n,\eta}$, with $g$ the Land\'e g-factor, $\mu_{B}$ the Bohr magneton, $\bm{B}_{n,\eta}$ the local magnetic field, and $\bm{\sigma}$ the vector of the spin Pauli matrices. The other vector of Pauli matrices, i.e., $\bm{\tau}=\tau_{x,y,z}$, acts in particle-hole Nambu space. In order to simulate the stray field of the STT-arrays, the FEA method is employed using COMSOL Multiphysics~\cite{comsol}. The calculated stray-field is illustrated in Fig.~\ref{Fig2}(b) and is related to the local fields $\bm{B}_{n,\eta}$ discussed above. 

In a similar fashion, we define the following Hamiltonian for the superconductor
\bea
H_{\rm sc}=
\frac{1}{2}\sum_{n}
\left[a_n^\dag{\cal H}_{\rm sc}^Aa_n+
\Big(a_n^\dag{\cal H}_{\rm sc}^Ba_{n+1}+
{\rm H.c.}\Big)\right],
\eea

\noi where we introduced the Hamiltonian matrices:
\bea
\mathcal{H}_{\rm sc}^A=(4t-\mu_{\rm sc})\tau_z+\Delta\tau_x\,\,\,{\rm and}\,\,\,
\mathcal{H}_{\rm sc}^B=-t\tau_z,\quad
\eea

\noi with $a_n$, $\mu_{\rm sc}$ and $\Delta$, correspondingly defining the BdG spinor, the chemical potential and the pair potential of the superconductor. Lastly, the coupling Hamiltonian is
\bea
H_{\rm c}=
\frac{1}{2}\sum_{n,\eta={\rm L,R}}
\Big(c_{n,\eta}^\dag
T_{\eta}a_n+
{\rm H.c.}\Big)\,\,\,{\rm with}
\,\,\,
T_{\eta}=-t_{\eta}\tau_z,\quad
\eea

\noi where $t_{\eta}$ represents the coupling between the $\eta$-channel and the adjacent sites in the superconductor.

Based on this model, we particularly focus on the CAR transmission coefficient $T_{\rm CAR}$. Using the Green function method~\cite{prb85.035402, prb94.165441, jpcm21.344204, prl103.167003}, we end up with the expression
\begin{equation}
T_{\rm CAR}=\frac{e^2}{h}\mathrm{Tr}\big(\Gamma_{ee}^{\rm L}\mathcal{G}_{eh}\Gamma_{hh}^{\rm R}\mathcal{G}_{he}^{\dag}\big),
\end{equation}

\noi where $e$ ($h$) indicates the electron (hole) degree of freedom in the Nambu space. The Green function of the whole system, $\mathcal{G}(E)$, is given by
\begin{equation}
\mathcal{G}(E)=\big(E-H_{\rm total}-\Sigma_{\rm L}-\Sigma_{\rm R}\big)^{-1},
\end{equation}

\noi where the retarded self-energies $\Sigma_{\eta={\rm L,R}}$ result from the coupling to the normal leads. The line-width function $\Gamma_{\eta}$ for the $\eta$-channel is given as $\Gamma_{\eta}=i(\Sigma_{\eta}-\Sigma_{\eta}^{\dagger})$. In our calculation we take the wide-band limit. Hence, each $\Gamma_{\eta}$ is an energy-independent constant. Similarly, we calculate the LAR transmission coefficient $T_{\rm LAR}$ which reads
\begin{equation}
T_{\rm LAR}=\frac{e^2}{h}\mathrm{Tr}\big(\Gamma_{ee}^{\rm L}\mathcal{G}_{eh}\Gamma_{hh}^{\rm L}\mathcal{G}_{he}^{\dagger}\big).
\end{equation}

\noi To transparently compare the amplitudes for CAR and LAR, we directly refer $T_{\rm CAR}$ and $T_{\rm LAR}$ as the correspon\-ding linear conductances $G_{\rm CAR}$ and $G_{\rm LAR}$. We define $\Delta G\equiv G_{\rm CAR}-G_{\rm LAR}$ and calculate $\Delta G$ versus $\mu_\eta$. The calculated $\Delta G$ in the presence of the STT-array is shown in Fig.~\ref{Fig2}(d). To further facilitate their comparison, we also include results in the absence of the STT-array in Fig.~\ref{Fig2}(c). 

From our results, it becomes evident that without the local stray-field generated by the STT-array, the electron transport is dominated by the LAR process. The CAR coefficient can only slightly exceed its LAR counterpart at a few resonant tunneling points. In stark contrast, the addition of the stray-field stemming from the FPTA rendered the CAR process dominant in a large region of the $\mu_{\rm L,R}$ parameter space. 

The CAR enhancement is attributable to the delicately-designed FPTA cells, and more specifically, to the parallel and antiparallel configurations dictating the STT-arrays. This design feature plays a twofold role. First, the stray-field establishes opposite Zeeman splittings for the two 1D-channels, and second, its oscillating part further mixes the spin-up and spin-down via the synthetic SOI. The spin degeneracy of the energy bands is further lifted and therefore favors the CAR. 
Our FPTA mechanism is clearly distinct from proposals relying on uniform magnetic fields or ferromagnets to mediate the CAR enhancement. In fact, since it mainly relies on the antiparallel electron spin orientation on the two channels, it further allows relaxing the requirement for chemical potential tuning. Moreover, the magnetic-field-induced Cooper-pair breaking effects can be here greatly mitigated due to the local nature of the stray field.

\begin{figure*}[ht]
\centering \includegraphics[width=17.5 cm]{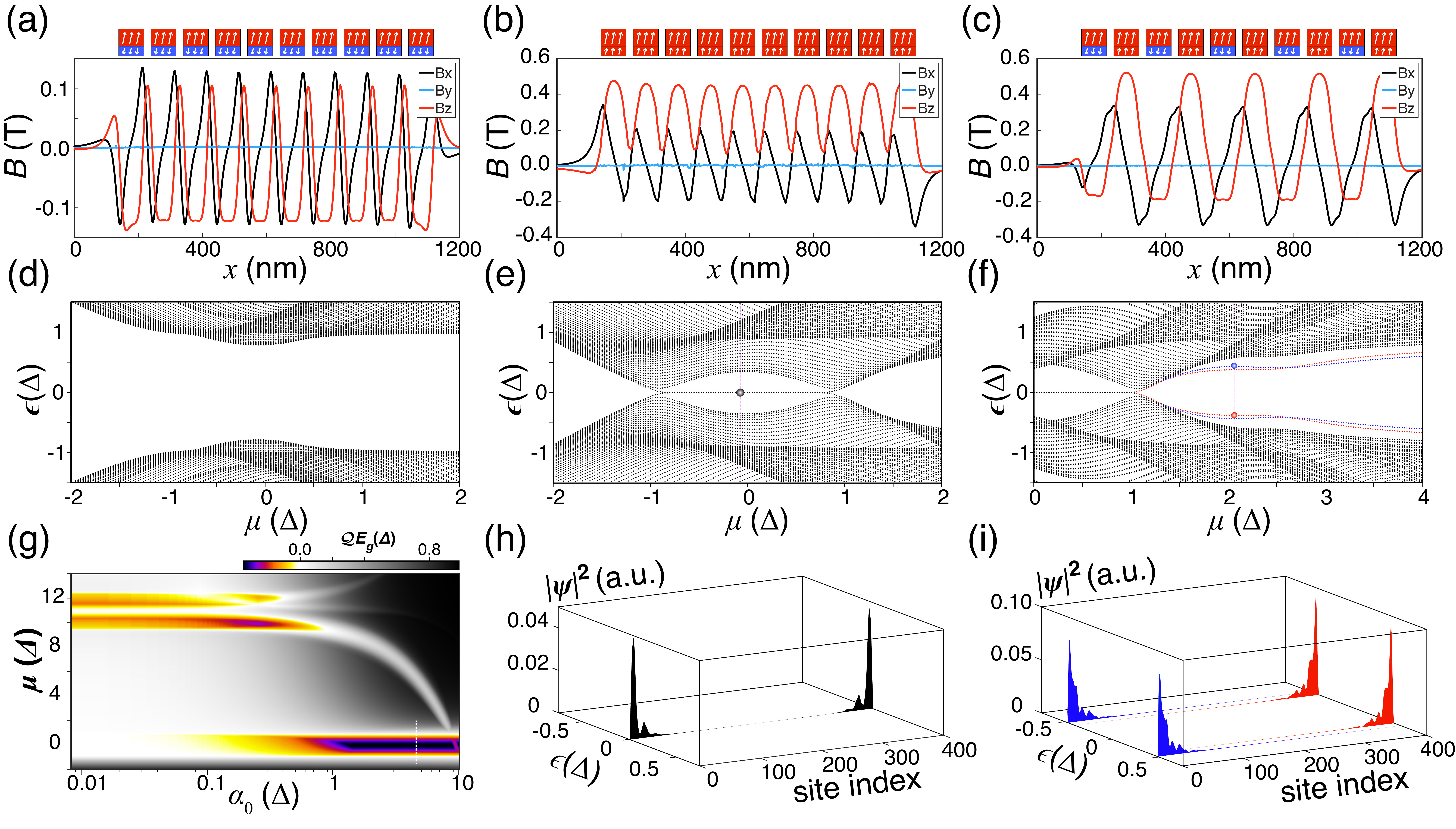}
\caption{\label{Fig3} Three typical designs and the corresponding energy spectra and wave functions. The three typical configurations and the generated magnetic fields distribution: (a) the antiparallel configuration, (b) the parallel configuration, and (c) the cross configuration. (d)-(f) Energy spectra vs. the chemical potential $\mu$ for the configurations in (a)-(c) respectively. (g) System energy gap, $E_g$, which is multiplied by the topological invariant, $\boldsymbol{\mathcal{Q}}$ (see definition in the text), as a function of the intrinsic SOI and chemical potential. The color-coded region indicates the topologically-nontrivial regime. In the presence of an oscillating stray-field, the system can be tuned into topological phase even with small intrinsic SOI. (h) Probability densities of the wave functions for the state highlighted in (e). The wave function illustrates a non-local property. (i) Probability densities of the wave functions for the state highlighted in (f). In the model calculation, the stray-field value are calculated using FEA-method, and the related parameters are $\Delta_{\rm ind}=200$~$\mu$eV, $t=28.6\Delta_{\rm ind}$, $\alpha_{0}=4.5\Delta_{\rm ind}$, $g=30$, $a=20$~nm, $m^*=0.015m_e$. The total site number is 400. The site number in one period is 5 for the antiparallel and the parallel configurations, and 10 for the cross configuration.} 
\end{figure*}

\subsection{FPTA-Controlled Topological Phase Transition}\label{sec:SectionIII_B}

In the second case-study, we demonstrate how to employ the STT-based FPTA to control the topological order appearing in a 1D SM-SC hybrid nanowire. 

The structure studied here is a 1D-semiconductor nanowire, which is either fabricated by means of direct epitaxy~\cite{Krogstrup2015, Vaitiekenas2018SAG}, or it is gate-defined using a 2DEG substrate~\cite{Shabani2016}. In addition, an array of STT-cells is arranged on top of the nanowire. The latter is also under the influen\-ce of a Rashba-type SOI, and the chemical potential of the nanowire can be tuned by a nearby-global gate electrode. A superconductor is further coupled to the nanowire. This setup is actually half of the system shown in Fig.~\ref{Fig1}(e) or Fig.~\ref{Fig2}(a).

The various configurations of the STT-cells give rise to a diversity of spatially-oscillating magnetic fields, which in turn control the topological properties of the target material, i.e., the hybrid nanowire in the present context. In Figs.~\ref{Fig3}(a)-(c), we depict the numerically-calculated stray-field distribution induced in the 1D-nanowire, for three different STT-arrangements, namely, the antiparallel-, the parallel-, and the cross-configuration. See the top panel schematics in Fig.~\ref{Fig3}. The STT-array generates an oscillating stray-field in the $xz$ spin plane, which can greatly modify the Zeeman energy and the effective SOI of the nanowire. Depending on the oscillation amplitude and wavelength, as well as the chemical potential seen by the nanowire, various topological phases can be achieved. 

To numerically infer the topological scenarios that become accessible here, we employ once again a tight-binding model combined with the FEA method. The BdG Hamiltonian of the hybrid nanowire is compactly represented as: ${\cal H}_{\rm hybrid-nanowire}={\cal H}_{\rm single-ch}+{\cal H}_{\rm sc}$, with ${\cal H}_{\rm single-ch}$ obtained from Eq.~\eqref{eq:Hch} after restricting to a single channel. The matrix part ${\cal H}_{\rm sc}=\Delta_{\rm ind}\tau_x$ contains the frequency-independent pairing gap $\Delta_{\rm ind}$ which emerges after integrating out the degrees of freedom of the superconductor~\cite{Potter and Lee}, which is dictated by a bulk gap $\Delta$. 

Another crucial ingredient which is considered here and was absent in the model of the previous section, is that we also assume the presence of a Rashba-type SOI oriented along the $y$ spin axis. This is effectively incorporated in our model via replacing $t$ by $\sqrt{t^2+\alpha_0^2}$ and considering the unitary transformation~\cite{BrauneckerSOC}
\bea
\bm{\mathcal{V}}_n\cdot \bm{\sigma}\mapsto
e^{-i\chi n\sigma_y/2}
\bm{\mathcal{V}}_n\cdot \bm{\sigma}e^{i\chi n\sigma_y/2}\,.
\eea
where $\tan \chi=\arctan (\alpha_0/t)$, with $\alpha_0$ denoting the SOI energy scale in our tight-binding model, which in the continuum model it corresponds to a SOI of a strength $\alpha\equiv 2a\alpha_0=36$ meV$\cdot$nm. From the above, it becomes apparent that the SOI mixes the $x$ and $z$ magnetic field components induced by the STT-cells, and tends to stabilize a spiraling spatial profile, since it enhances the winding density of $\bm{\mathcal{V}}_n$, i.e.:
\bea 
w_n=\big(\bm{\mathcal{V}}_{n+1}\times \bm{\mathcal{V}}_n\big)_y\,.
\eea

\begin{figure*}[ht]
\centering \includegraphics[width=17 cm]{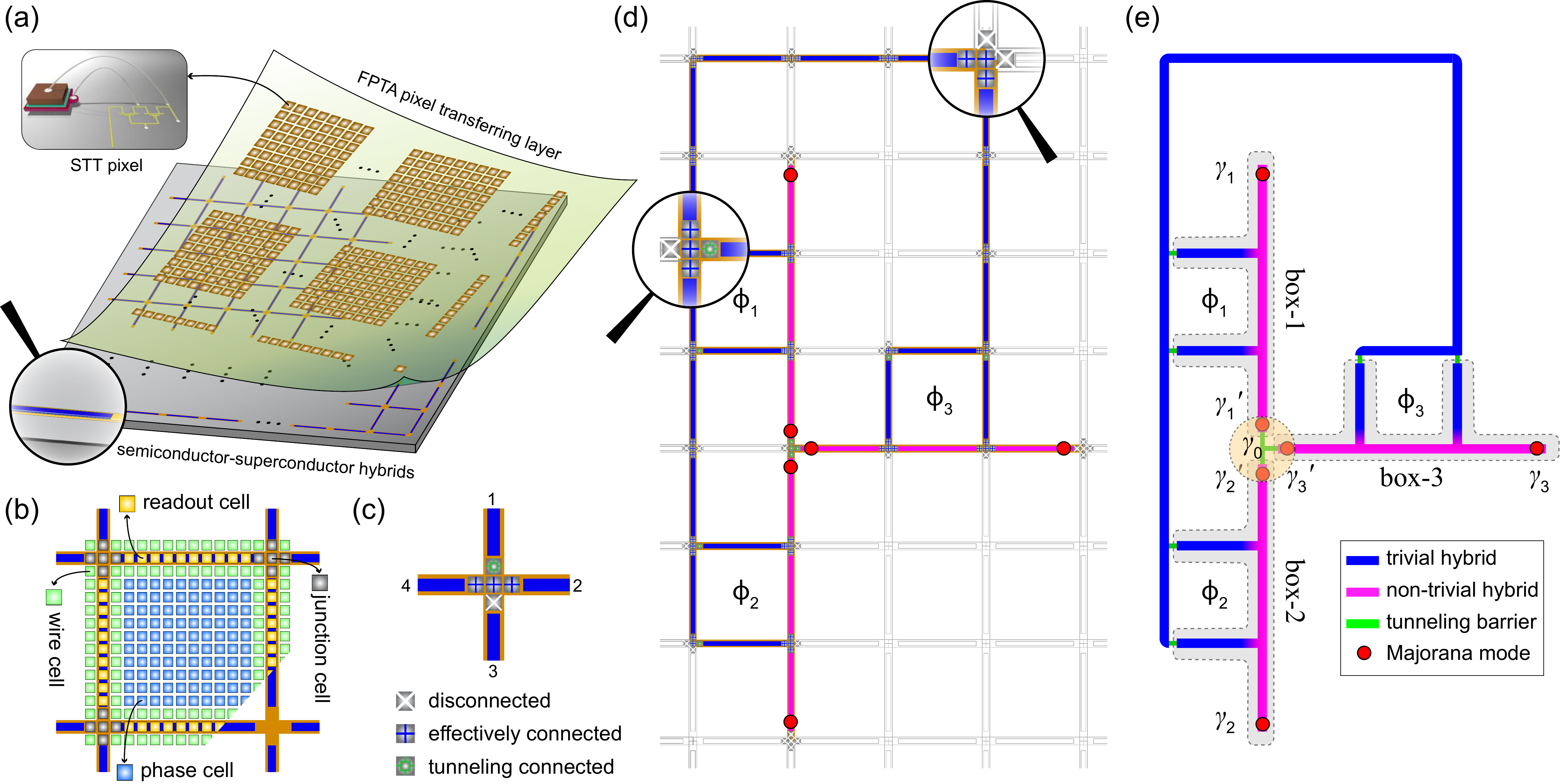}
\caption{\label{Fig4} A SM-SC hybrid network interfaced with an FPTA. (a) Schematic of a SM-SC hybrid network with a separated FPTA layer. The FPTA layer can be transferred on and align with the target network substrate via the van der Waals integration method. (b) Cell function assignment in a unit grid cell of the network. Wire cells are used to set the topologically (non)trivial phase of the hybrid segment; phase cells pin the phase of the SQUID in the unit cell; junction cells are used to control the electrical transparency of the junction; and readout cells are employed for the dispersive readout of the quantum capacitance of the box. (c) Junction cell configurations in different operation modes. In the ``disconnected" regime, the Cooper pair and single quasiparticle transport channels are both completely pinched off, while the ``effectively connected" regime allows for a sizeable Josephson current to flow through. A junction cell operating in the ``tunneling" mode allows us to set an optimal Josephson energy for the braiding operation. (d) An FPTA configuration suitable for the implementation of a Coulomb \& flux-assisted braiding protocol~\cite{Heck2012, Hyart2013}. (e) Equivalent schematic to (d).} 
\end{figure*}
%---------------------------------------------------------------------------------------------------------------------------------
%%%%%%%%%%%%%%%%%%%%%%%%%%%%%%%%%%%%%%%%%%%%%%%%%%%%%

As reflected in the results of our numerical calculations, which are depicted in Figs.~\ref{Fig3}(d)-(f), the energy spectra obtained from the above model for the three distinct configurations are clearly different. Below we discuss our results for each configuration.

\bt{\textit{Antiparallel Configuration.}} When the STT-cell array is in the antiparallel configuration, the energy spectrum shows that the system is in a trivial gapped superconducting phase. In this configuration, the stray-field from the STT free-layer largely counteracts with the fix-layer stray field, and the net magnetic field is not sufficiently strong to compensate for the s-wave pairing gap.

\bt{\textit{Parallel Configuration.}} If the free-layers of the STT-cells are all switched to the spin-polarizations that are parallel to the fix-layers, the system is then set to the parallel configuration. The net magnetic field is much higher than the antiparallel configuration. As shown in Fig.~\ref{Fig3}(e), the gap in the bulk energy spectrum closes and reopens as a function of the chemical potential, thus effecting a transition from a topologically trivial to a nontrivial phase, which is accompanied by the emergence of zero-energy states. The topological indicator of the phase can be seen from the topological invariant $\boldsymbol{\mathcal{Q}}$, which is defined as:
\begin{equation}
\boldsymbol{\mathcal{Q}}=\mathrm{sgn}[\mathrm{Pf}\,\mathcal{A}(0)]\,\mathrm{sgn}[\mathrm{Pf}\,\mathcal{A}(\pi/a)],
\end{equation} where $\mathrm{sgn}$ is the sign function, $\mathrm{Pf}$ is the Pfaffian function, and $\mathcal{A}$ is the target Hamiltonian matrix written in the Majorana basis with a periodic boundary~\cite{pu44.131}. In practice, $\boldsymbol{\mathcal{Q}}$ is calculated by using the routine in Ref.~\onlinecite{acmtms38.30}. 
By inspecting the wave functions of these zero-energy states [Fig.~\ref{Fig3}(h)], we verify they indeed possess a nonlocal character and describe MZMs. 

It is well-established that when a 1D-superconducting nanowire feels the presence of a strong SOI-field together with an orthogonal magnetic field to it, the nanowire becomes a topological superconductor (TSC) with a pair of MZMs appearing at the nanowire ends~\cite{Lutchyn2010, Oreg2010}. In contrast to the global Zeeman field considered in the standard nanowire MZM proposals~\cite{Lutchyn2010,Oreg2010}, the local stray-field from the STT-cells contains both a uniform component and a spiral (or generally oscillating) component. The spiral component provides a strong synthetic SOI and thus exempts the material from the requirement for intrinsic SOI~\cite{BrauneckerSOC, Kjaergaard2012}. Given the stray-field value from the FEA simulation, the phase diagram in Fig.~\ref{Fig3}(g) shows that the system can maintain a topological phase even for a near-zero inherent SOI. It is noted that, to highlight the topological phase, we integrate the topological invariant $\boldsymbol{\mathcal{Q}}$ into the colorized phase diagram.

\bt{\textit{Cross configuration.}} When the spin-polarization of the free-layers are configured to align in parallel and antiparallel to the fix-layers in an alternating fashion, the setup is in a cross-type configuration. In this case, there are two phases in the spectrum, separated by a gap clos\-ing point. In the regime $\mu<0.27\Delta_{\rm ind}$, the MZMs persist. Instead, in the regime $\mu>0.27\Delta_{\rm ind}$, we find Andreev modes which appear energetically near the bulk-gap edge. These are depicted in Fig.~\ref{Fig3}(i). In the absence of a pairing gap, these Andreev bound states (ABSs) would correspond to localized electronic excitations carrying fractional $e/2$ charge, similar to the ones encountered in the Jackiw-Rebbi and Su-Schrieffer-Heeger models~\cite{Jackiw1976, SSH1979}. The formation of the FF-phase arises here because the period of the spiral component of the stray-field is set to $2\pi/4k_{so}$, where $1/k_{so}=\hbar/m_e^*\alpha$ defines the SOI length. As a result, an extra coupling appears between the $k=\pm2k_{so}$ points of the exterior dispersion parabola branches of the Rashba nanowire~\cite{Klinovaja2012, Rainis2014}. 
To realize the FF-phase, the amplitude, period, and phase of the oscillating stray-field have to satisfy certain conditions. The STT-cell allows the direct control of both the amplitude and the period of the oscillating field, while the phase tuning can be effected by depleting part of the nanowire within an appro\-pria\-te length. For the latter, one can rely on the electric-potential control mode of the STT-cell.

Concluding this case-study, we point out that we have numerically demonstrated that the FPTA provides a control knob for effecting a variety of topological phase transitions in the hybrid nanowire, which span the tri\-vial superconducting phase, the MZM-phase, and the FF-phase. 

%============================###########===========================
%============================###########===========================

\begin{table*}[t!]
\centering
\caption{Majorana braiding protocol in the hybrid network - FPTA coupled system. The protocol starts with the generation of three pairs of MZMs, with the three inner MZMs at the trijunction hybridizing into one single MZM, as depicted in Fig.~\ref{Fig4}(e). Braiding is realized through the flux control of the MZM couplings in a given box, and specifically by sequentially switching these on/off in a cyclic manner. The readout operation is not described here.}
\label{table-1}
\begin{tabular}{@{}c|l|l|l@{}}
\toprule
\hline
\hline
Step & Operating cells & STT Operating Mode & Operation Details \\ \midrule
\hline
\hline
 1 & wire cell & voltage control\footnote{The wire cells can also operate in the spin-polarization control mode at this step.} &  generate MZM pairs $\gamma_s$/$\gamma_s^\prime$, with $s= 1,2,3$ \\
 2 & junction cell & voltage control & defining Majorana box-$s$; fine tune  $E_{{\rm C},s}/E_{{\rm J},s}, s=1,2,3$\\
 3 & junction cell & voltage control & fuse the three inner Majoranas $\gamma_s^\prime, s= 1,2,3$ into $\gamma_{0}$\\
 4 & phase cell & spin-polarization control & initialize phases $\Phi_1 \Rightarrow \Phi_{\rm min,1}$, $\Phi_2 \Rightarrow \Phi_{\rm min,2}$, $\Phi_3 \Rightarrow \Phi_{\rm max,3}$\\
 5 & phase cell & spin-polarization control & phase tuning $\Phi_1 \Rightarrow \Phi_{\max,1}$\\
 6 & phase cell & spin-polarization control & phase tuning $\Phi_3 \Rightarrow \Phi_{\min,3}$\\
 7 & phase cell & spin-polarization control & phase tuning $\Phi_2 \Rightarrow\Phi_{\max,2}$\\
 8 & phase cell & spin-polarization control & phase tuning $\Phi_1 \Rightarrow \Phi_{\min,1}$\\
 9 & phase cell & spin-polarization control & phase tuning $\Phi_3 \Rightarrow \Phi_{\max,3}$\\
 10 & phase cell & spin-polarization control & phase tuning $\Phi_2 \Rightarrow \Phi_{\min,2}$ \\
 11 & phase cell & spin-polarization control & phase tuning $\Phi_3 \Rightarrow \Phi_{\min,3}$\\
\hline
\hline
\end{tabular}
\label{table:Cells}
\end{table*}

\subsection{FPTA-Based Braiding in a Hybrid Network}\label{sec:SectionIII_C}

In the previous section, we demonstrated how the STT-based FPTA provides a flexible control knob over electron transport and the topological phase transitions occurring in low-dimensional hybrid quantum systems. In the present case-study, we elaborate on FPTA-based protocols that enable us to perform non-Abelian braiding using MZMs.

For this purpose, we design a system composed of a SM-SC hybrid network substrate~\cite{Krizek2018, Vaitiekenas2018SAG} and a separated STT-FPTA layer, as shown in Fig.~\ref{Fig4}(a). Depending on the situation, the hybrid network and the FPTA layer can be assembled using the van der Waals integration method~\cite{Liu2019}, which can simplify the material engi\-nee\-ring process and minimize the possible interface disorders. The considered network has a grid-like structure, and there exist four SM-SC hybrid sections in each square unit. Moreover, the electron density is fully-controllable at each cross. For details see Figs.~\ref{Fig4}(b)-(c). Ensuring a good alignment of the FPTA "stamped" on top of the hybrid grid layer endows the latter with a high degree of multi-functionality. The advantage of this platform is that it is compatible with a diversity of braiding and other quantum information protocols, independently on whether these involve: (i) the exchange of MZMs in the physical space~\cite{Alicea2011,Sau2011}, or (ii) the coupling of MZMs using charging effects~\cite{Heck2012, Hyart2013,Aasen2016}, or (iii) projective measurements on the MZM-defined qubits~\cite{Karzig2017}. 

Out of the above possibilities, we adopt the Coulomb \& flux-assisted protocol~\cite{Heck2012, Hyart2013} to demonstrate an FPTA-driven braiding process. To carry out this task, it is necessary to separate the FPTA cells into four different types, i.e., wire, junction, phase, and readout cells. For a sketch see Fig.~\ref{Fig4}(b). Below and in Table~\ref{table:Cells}, we specify the functionalities and operation modes of each cell type.

\begin{itemize}[leftmargin=*]

\item\bt{Wire Cell.} The wire cells correspond to the FPTA basic units which are distributed along the SM-SC hybrid sections and feature a green color-coding in Fig.~\ref{Fig4}(b). Similar to the case-study in Sec.~\ref{sec:SectionIII_B}, the wire cells selects whether the hybrid segment lies in the topologically-trivial superconductor phase or the nontrivial phase that is effectively equivalent to a spinless $p$-wave superconductor. This is achieved by varying the chemical potential (voltage-control mode) and/or the spiral field oscillations (spin-polarization mode). The TSC segments host MZMs which are involved in brai\-ding, while the trivial superconductors constitute the grounding bus.

\item\bt{Junction Cell.} These FPTA-units are located in the uncovered semiconductor cross area. The junction cells control the direction of the supercurrent flow, define the Cooper pair or Majorana box, and adjust the ratio of $E_C/E_J$, which denotes the charging/Josephson energy of the junction. As shown in Fig.~\ref{Fig4}(c), the junction cells control the strength of the coupling appea\-ring between hybrid segments that meet at it. The latter can be fully-disconnected or become coupled under conditions of moderate or perfect transparency.

\item\bt{Phase Cell.} The FPTA-units appear in the central area of the grid and are shown with blue squares in Fig.~\ref{Fig4}(b). These mainly tune the superconducting phase across the hybrid junction.

\item\bt{Readout Cell.} Majorana parity measurements can be performed by parity-to-charge conversion and electric gate dispersive readout technology~\cite{Colless2013} through the readout cells. The readout cells are arranged along the superconductor. See the yellow squares in Fig.~\ref{Fig4}(b).
\end{itemize}

We remark that the only difference among the four types of FPTA cells is that these are located at dif\-fe\-rent positions relative to the hybrid network. The cell structures themselves are otherwise identical.

In the above structure, three pairs of MZMs can be created and form a Majorana tri-junction as in Figs.~\ref{Fig4}(d)-(e). The low-energy Hamiltonian of the tri-junction consists of two parts, i.e., (i) the Hamiltonian $\mathcal H_T$ which describes the coupling of the MZMs, and originates from electron tunneling across the junction~\cite{Alicea2011}, and, (ii) the Hamiltonian $\mathcal H_C$ which contains the coupling between MZMs of a given ``Majorana-box", which has a capacitive cha\-rac\-ter~\cite{FuTeleportation}, and results from the charging energy of the island~\cite{Heck2012,Hyart2013}. 

The tunneling-related Hamiltonian term reads:
\begin{equation}
    \mathcal H_T=iE_{M}\left(\cos{\theta_{12}}\gamma_1'\gamma_2'+\cos{\theta_{23}}\gamma_2'\gamma_3'+\cos{\theta_{31}}\gamma_3'\gamma_1'\right),
\end{equation}

\noi with $E_M$ the energy scale dictating the MZM coupling and $\theta_{ss'}$ the gauge-invariant phase difference between the neighboring boxes $s$ and $s'$. On the other hand, Coulomb-coupling term is represented as:
\bea
\mathcal H_C=
-i\sum_{s=1}^3U_s\gamma_s\gamma_s',
\eea

\noi in which $U_s$ is the coupling between the two remote MZMs harbored in box-$s$, due to non-negligible char\-ging effects. The strength of this type of coupling is given by~\cite{Heck2012}:
\begin{equation}
U_s=16\left(\frac{E_{C,s}E_{J,s}^3}{2\pi^2}\right)^{1/4}e^{-\sqrt{8E_{J,s}/E_{C,s}}}\cos(\pi Q_s/e),
\label{eq:Uk}
\end{equation}

\noi with $E_{J,s}=2E_{0,s}\cos (\pi \Phi/\Phi_0)$ the Josephson energy of box-$s$ and $Q_s$ the induced charge offset from nearby electrode gates. As first shown in Ref.~\onlinecite{Heck2012}, substituting $E_{J,s}$ into $U_s$ in Eq.~\ref{eq:Uk}, yields $U_s\appropto {\rm Exp}[-4\sqrt{(E_{0,s}/E_{C,s})\cos(\pi \Phi_s/\Phi_0)}]$, i.e., the Coulomb-coupling strongly depends on the flux $\Phi_s$. Therefore, the flux can serve as a switch for the MZM coupling of box-$s$, and the flux/phase can be controlled by the phase cells of the FPTA. We denote the fluxes corresponding to $U_{\max,s}$ and $U_{\min,s}$ as $\Phi_{\max,s}$ and $\Phi_{\min,s}$, respectively.

The full protocol of braiding a pair of MZMs in the network is shown in Table~\ref{table:Cells}, which is in line with the sequence proposed in Ref.~\onlinecite{Heck2012}. This method accounts for both, i.e., the steps required to generate the MZMs, as well as the and knob manipulations that effect the desired Majorana braiding. 

Concluding, we remark once again that the FPTA approach can be harnessed to realize alternative braiding or quantum computing operations. For instance, one can employ the above-discussed FPTA design to also implement a gate-controlled braiding protocol~\cite{Aasen2016}. 

\subsection{FPTA-Engineering of Synthetic Weyl Points}\label{sec:SectionIII_D}

The advantageous aspects of scalability and the high degree of spatiotemporal control over the various phy\-si\-cal quantities that become accessible in the FPTA framework, provide a unique playground for ABS engineering and the observation of novel quantized transport phenomena~\cite{Meyer, LevchenkoTop, PK_PRL_2019, MTM_PRB_2019, Balseiro, Sakurai, Houzet} in multi-terminal Josephson junctions, similar to the ones recently experimentally fabricated in Ref.~\onlinecite{Manucharyan}. The phenomena in question have a topological origin and can be attributed to the emergence of Weyl points in the synthetic ABS band structure.

The scope of this case-study is to unveil how to crea\-te such synthetic Weyl points and thus unlock the concomitant topological transport within the FPTA framework. 

\begin{figure}[t]
\centering
\includegraphics[width=1\columnwidth]{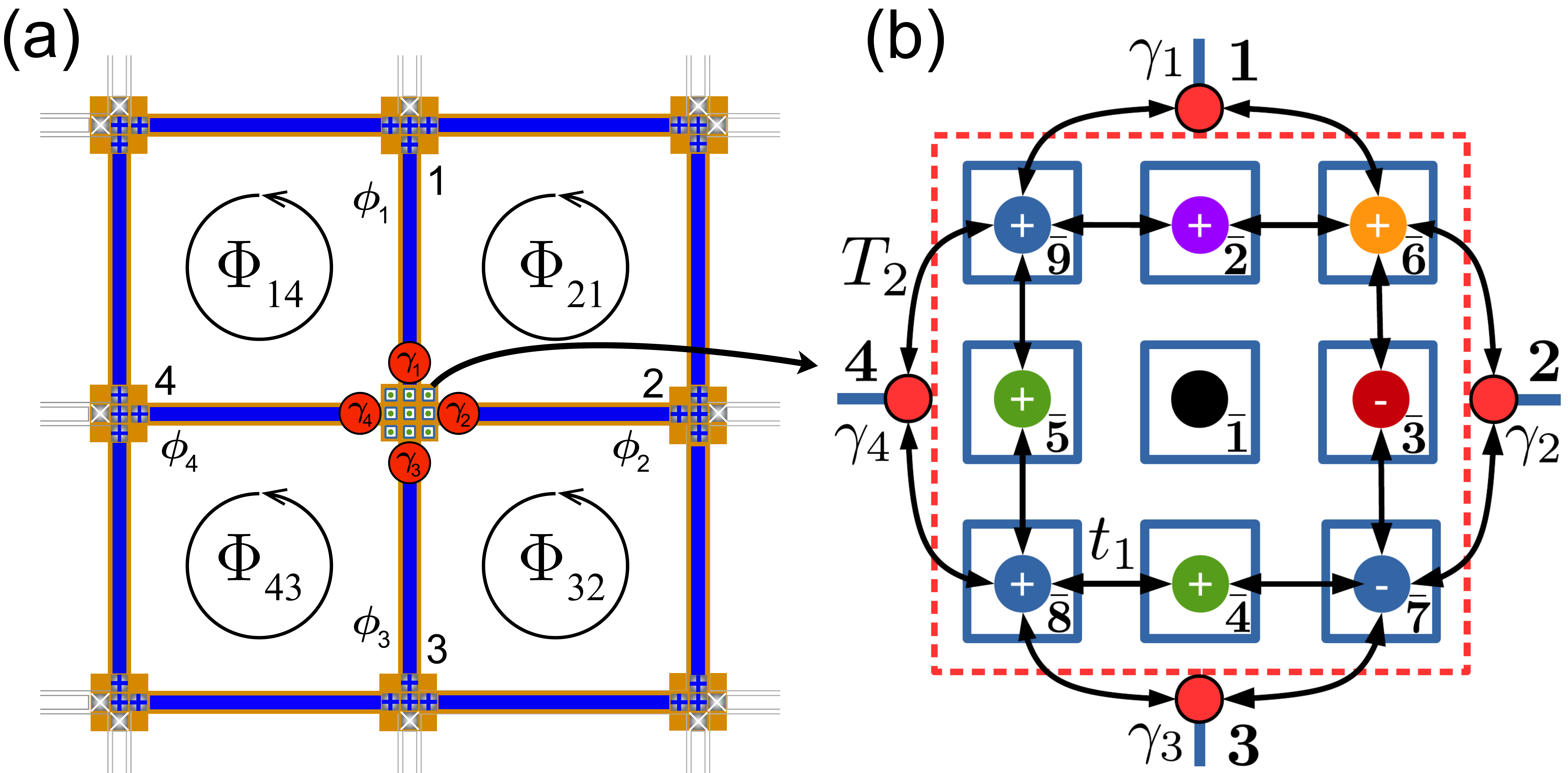}
\caption{(a) Multi-terminal Josephson junction consisting of four topological nanowires with each one harboring a single MZM per edge ($\gamma_{1,2,3,4}$). It is similar to the structure in Fig.~\ref{Fig4}, but only $3\times 3$ junction cells are illustrated here. Four fluxes $\Phi_{ss'}$ pierce the four loops formed by the hybrid segments. In the weak tunnel-coupling regime, the properties of the junction can be described by the four coupled MZMs. (b) The proposed middle junction (MJ) configuration for engineering Weyl points in synthetic space, which allow for quantized transport phenomena. Here the phase differences $(\phi_{21},\phi_{43})$ play the role of the pumping va\-ria\-bles $(\phi,\theta)$ driving the topological transport. Regarding the remaining parameters, we consider $\phi_{14}=\phi_{32}=\pi$, $1/|U_{\bar{1}}|=0$, $U_{\bar{2},\bar{4},\bar{5}}>0$, $U_{\bar{3}}<0$, $U_{\bar{4}}=U_{\bar{5}}$, and $U_{\bar{7}}=-U_{\bar{8},\bar{9}}<0$. With $\pm$, we denote the sign of the potential energy at each STT.}
\label{Fig5}
\end{figure}

The system studied here is similar to the structure shown in Fig.~\ref{Fig4}. We restrict to junctions construed by four nanowires in the topologically-nontrivial regime, each one of which harbors a single MZM per edge. See Fig.~\ref{Fig5}. The middle junction (MJ) consists of nine FPTA-units, which modify the electrostatic environment of the 2DEG lying underneath them by generating a nontrivial landscape of potential energy $U_{\bar{n}}$, where $\bar{n}=\bar{1},\bar{2},...,\bar{9}$ denotes sites within the MJ area. As shown in Fig.~\ref{Fig5}(a), the MJ is surrounded by four grids which are defined for pairs of nanowires labeled by $s,s'=1,2,3,4$. Each grid encircles a respective flux $\Phi_{ss'}$ induced by phase-cells.

We employ the following Hamiltonian for the 2DEG underneath the MJ:
\begin{align}
{\cal H}_{\rm MJ}=\sum_{\sigma=\uparrow,\downarrow}\sum_{\bar{n},\bar{m}}c_{\bar{n}\sigma}^{\dag}\left(t_{\bar{n}\bar{m}}e^{i\pi \int_{\bm{R}_{\bar{n}}}^{\bm{R}_{\bar{m}}}
\bm{A}\cdot{\rm d}\bm{r}/\Phi_0}+U_{\bar{n}}\delta_{\bar{n}\bar{m}}\right)c_{\bar{m}\sigma},
\label{eq:MJ Hamiltonian}
\end{align}

\noi with the indices $\bar{n}$ and $\bar{m}$ being restricted to the nine sites buried under the corresponding STTs. In the above, $t_{\tilde{n}\tilde{m}}\in\mathbb{R}$ are generally nonzero only for the nearest ($t_1$) and the next-nearest neighbours ($t_2$). By means of sui\-ta\-ble so-called Peierls phase factors we further included the line integrals of the vector potential $\bm{A}$, that allow de\-scri\-bing the possible induction of flux $\Phi_{\rm MJ}$ through the MJ when all four superconducting phase differences are fixed to be nonzero. Specifically, the constraint 
\bea
\Phi_{\rm MJ}=I_{\rm Grid}
+\left(\ell-\frac{\phi_{14}+\phi_{43}+\phi_{32}+\phi_{21}}{2\pi}\right)\Phi_0\quad
\eea

\noi applies, where $\Phi_0=h/(2e)$ denotes the flux quantum, $\ell\in\mathbb{Z}$, and $I_{\rm Grid}$ (here in units of flux) corresponds to the current circulating in the outer grid which is formed by the four square sub-grids. In the above, we introduced the superconducting phase differences $\phi_{ss'}=\phi_s-\phi_{s'}=2\pi\Phi_{ss'}/\Phi_0$, with $\phi_s,\phi_s'$ ($s,s'=1,2,3,4$) the superconducting phase of each TSC. The above relation is compatible with the fluxoid-quantization condition~\cite{LittleParks1962}: $\Phi_{14}+\Phi_{43}+\Phi_{32}+\Phi_{21}+\Phi_{\rm MJ}=I_{\rm Grid}+
\ell\Phi_0$. One distinguishes two limiting scenarios. In the first (second), the superconducting elements fully (do not) screen the magnetic field, thus implying $I_{\rm Grid}=0$ ($\Phi_{\rm MJ}=0$), since no external magnetic field is applied in the MJ. The integer $\ell$ denotes the vorticity of the flu\-xoid which develops in order to minimize the energy of the system~\cite{Liu2001,Vaitiekenas2020}.
Both limiting scenarios lead to qualitatively similar effects when it comes to engineering synthetic Weyl points. In the remainder of this paragraph, we focus on the first situation, where $I_{\rm Grid}=0$, since it leads to a richer behavior. The other possibility is briefly discussed in Appendix~\ref{sec:AppendixB}. To proceed, we additionally set $\ell=0$, since one can either tune the device parameters so that this value is energetically favored, or focus in the vicinity of a suitable Weyl point that is dictated by $\ell=0$. 

In the weak-tunneling and thus low-energy regime, the resulting ABS spectrum and the related Weyl points are mainly determined by the MZMs which become effectively coupled due to the inter-nanowire tunnel couplings. From the analysis presented in Appendix~\ref{sec:AppendixB}, we find that the Hamiltonian de\-scri\-bing the coupling between the four MZMs, and that governs the ABS spectrum of the junction, takes the form~\cite{PK_PRL_2019} ${\cal H}_{\rm ABS}=\frac{i}{2}\bm{\Gamma}^{\intercal}\hat{B}\bm{\Gamma}$, where we rewrite the Hamiltonian in terms of the Majorana multi-component ope\-ra\-tor $\bm{\Gamma}^{\intercal}=(\gamma_1\,\,\gamma_2\,\,\gamma_3\,\,\gamma_4)$, its transpose $\bm{\Gamma}$, and the skew-symmetric matrix $\hat{B}$ given as:
\bea
\hat{B}_{ss'}=\sum_{n,m}{\rm Im}\Big({\rm T}_{nm}^{ss'}e^{i\phi_{ss'}/2}\Big)\,,\label{eq:Bees}
\eea

\noi where ${\rm T}_{nm}^{ss'}$ denote the matrix elements for electron tunneling involving the lattice sites $n$ and $m$ of the $s$-th and $s'$-th nanowire, respectively. The analysis below can be further simplified by separating matrix $\hat{B}$ into two sets labelled by $\bm{g}_1$ and $\bm{g}_2$, where $\bm{g}_1=-\big(B_{14}-B_{23},\,B_{13}+B_{24},\,B_{12}-B_{34}\big)$ and $\bm{g}_2=-\big(B_{14}+B_{23},\,B_{12}+B_{34},\,B_{13}-B_{24}\big)$. These are given in Appendix~\ref{sec:AppendixB}.

The coupling of the four MZMs results into four ABS dispersions with energies $\pm\varepsilon_\pm$. In the case of a full gap and under the assumption $|\bm{g}_2|\geq|\bm{g}_1|>0$, these obtain the form $\pm\varepsilon_\nu=\pm(|\bm{g}_2|+\nu|\bm{g}_1|)/2$. Weyl points emerge in the ABS spectrum when $|\bm{g}_1|=0$ is satisfied. Thus, the tou\-ching $\varepsilon_+=\varepsilon_-$ occurs when the three conditions $\{B_{14}=B_{23},\,B_{12}=B_{34},\,B_{13}=-B_{24}\}$ are simultaneously met. These can be experimentally implemented by tu\-ning the va\-rious tunnel couplings and/or fluxes.

\begin{figure}[t!]
\centering
\includegraphics[width=\columnwidth]{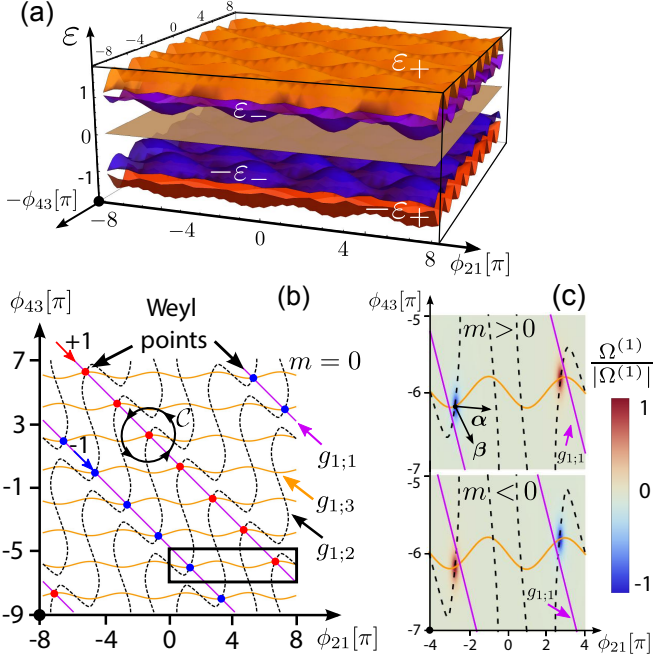}
\caption{(a) Fully-gapped ABS energy dispersions $\pm\varepsilon_{\pm}$ in units of $T_2^2/U_{\bar{9}}$ for $U_{\bar{2}}=-5U_{\bar{3}}$. The dispersions are $16\pi$-periodic in either phase difference $\phi_{21,43}$, due to the induced flux in the MJ area. (b) We depict $g_{1;1}(\phi_{21},\phi_{43})$ (magenta solid line), $g_{1;2}(\phi_{21},\phi_{43})$ (black dashed line) and $g_{1;3}(\phi_{21},\phi_{43})$ (orange solid line) overlaid on top of a heat map sketch of the sign of $\Omega^{(1)}_{\phi_{21},\phi_{43}}(\phi_{21},\phi_{43})$ for $m=0$. A number of $16$ Weyl points with charge $\pm1$ are obtained in the synthetic 3D space $(\phi_{21},\phi_{43},m)$ with $m-U_{\bar{2}}/U_{\bar{3}}\approx 1.52$ under the assumptions $4t_1^2=0.1U_{\bar{2}}U_{\bar{9}}$, $U_{\bar{2}}=0.1U_{\bar{4}}$, $U_{\bar{9}}=0.3U_{\bar{6}}$. The Weyl points located along a given line where $g_{1;1}=0$ possess the same topological charge. (c) Berry curvature $\Omega_{\phi_{21},\phi_{43}}^{(1)}(\phi_{21},\phi_{43})$ in a region enclosing two Weyl points, such as the ones enclosed in the frame shown in (b). In the limit $m=\pm0^+$, the Berry curvature is concentrated near the concomitant Dirac points in 2D synthetic space $(\phi_{21},\phi_{43})$, which become induced by the Weyl points.}
\label{Fig6}
\end{figure}

We now proceed with a concrete example, and consider: $T_1=t_2=1/|U_{\bar{1}}|=0$, $-U_{\bar{3}},U_{\bar{4}}=U_{\bar{5}}>0$, $U_{\bar{8}}=U_{\bar{9}}=-U_{\bar{7}}>0$, and $\phi_{14}=\phi_{32}=\pi$. By pinning additional values for the ratios of the potentials, we obtain a fully-gapped ABS spectrum as the one depicted in Fig.~\ref{Fig6}(a), with the energy expressed in units of $T_2^2/U_{\bar{9}}$, where $T_2$ denotes the strength for FPTA-mediated electron tunneling between nearest neighbor nanowires.

As shown in Fig.~\ref{Fig6}(b), the specific choice of parameter values leads to 16 Weyl points in the synthetic space $(\phi_{21},\phi_{43},m)$, with the synthetic plane $(\phi_{21},\phi_{43})$ defined in the reduced Brillouin zone $[-8\pi,8\pi)\times[-9\pi,7\pi)$. Note that the third synthetic dimension, i.e. $m$, defines also a mass term of a 2D Dirac Hamiltonian in the synthetic plane $(\phi_{21},\phi_{43})$, when it is instead viewed as a mere pa\-ra\-me\-ter. Given the parameter values and relations assumed, the mass term is approximately determined through the relation $m-U_{\bar{2}}/U_{\bar{3}}\approx1.52$. The monopole charge of each Weyl point is given by ${\rm sgn}\big[\bm{g}_1\cdot\big(\partial_{{\phi}_{21}}\bm{g}_1\times\partial_{{\phi}_{43}}\bm{g}_1)\big]$, with $\bm{g}_1$ evaluated at the Weyl point. Here, all the Weyl points possess a topological charge $\pm1$, with the sign determined by the direction of the vector $m(\bm{\alpha}\times\bm{\beta})$ with the velocity vectors $\bm{\alpha}$ and $\bm{\beta}$ depicted in Fig.~\ref{Fig6}(c). 

The two panels in Fig.~\ref{Fig6}(c) depict the sign of the Berry curvature~\cite{Berry,Niu} across the topological phase transition oc\-cu\-ring as we move from $m<0$ to $m>0$. For $m=\pm0^+$ the Berry curvature is concentrated in the vicinity of the Weyl point and yields a contribution of $\pm1/2$. We note that for a positive (negative) mass term, the sign of the resulting Berry curvature is the same as (is opposite to) the sign of the Weyl point's charge. Hence, we verify that the sign change of the Berry curvature across this transition, is equal to the respective Weyl charge. 

The overall sign-changing pattern of the Weyl points' charge further implies that when $m\neq0$, the integral of the Berry curvature $\Omega_{\phi_{21},\phi_{43}}^{(1)}(\phi_{21},\phi_{43})$ over the torus $[-8\pi,8\pi)^2$ is zero. None\-theless, as it has been pointed out in Ref.~\onlinecite{PK_PRL_2019}, since we are here dealing with synthetic topology, it is in principle possible to experimentally isolate the fractional contribution $\pm1/2$ to the Berry curvature, stemming from a single Weyl point. For this purpose, it is desirable to expe\-ri\-men\-tally rea\-li\-ze the limit $m\rightarrow0$, in which, the Berry curvature $\Omega_{\phi_{21},\phi_{43}}^{(1)}(\phi_{21},\phi_{43})$ becomes singular, and thus it is nonzero only at the Dirac points induced by the Weyl points in the $(\phi_{21},\phi_{43})$ plane. Since in this limit the Berry curvature is zero away from the Dirac points, it may appear experimentally challen\-ging to induce the FP by imple\-men\-ting the selective area swee\-ping principle, since this requires a very high resolution in the phase-difference variations. 

Given the above possible shortcomings, it appears that the most suited measurement stra\-te\-gy for isolating the contribution of a single Weyl point and inducing a topological $1e$-charge transfer, is by means of considering a pumping cycle which traces a closed path ${\cal C}$ in the synthetic $(\phi_{21},\phi_{43})$ plane that encloses the desired Dirac point. For an illustration, see Fig.~\ref{Fig6}(a). The resulting pumped charge is $\Delta Q_{\cal C}=e\upsilon$, with $\upsilon$ de\-no\-ting the vor\-ti\-ci\-ty of the Dirac point enclosed by ${\cal C}$. We find $\pi\upsilon=\varointctrclockwise_{\cal C}\Big[d\phi_{21}{\cal A}_{\phi_{21}}^{(1)}+d\phi_{43}{\cal A}_{\phi_{43}}^{(1)}\Big]\label{eq:vorticity}$, where $\bm{{\cal A}}^{(1)}=({\cal A}_{\phi_{21}}^{(1)},{\cal A}_{\phi_{43}}^{(1)})$ is the Berry vector potential defined for the $\bm{g}_1$ vector in the limit $m\rightarrow0$. Therefore, the presence of the under\-lying Weyl point can be reflected in the $\pi$-Berry phase, i.e. $\pi\upsilon$, picked up for a closed loop in synthetic space which encloses the respective vortex. 

\section{Conclusion}\label{sec:SectionIV}

To conclude, we have demonstrated how the FPTA can be employed to flexibly control electron transport and topological phase transitions, operate Majorana braiding, and engineer synthetic ABS dispersions containing Weyl points. Beyond these case-studies, FPTA shows the potential of engineering topological order by dres\-sing the parent-material with control layers. In this manner, the FPTA platform promises to unveil new physical phe\-no\-me\-na and utilities for applications. For instance, vortex-driving could be implemented in a controlled way when combining FPTA with metallic topological superconductors, thus allowing for non-Abelian braiding to be performed much easier than using scanning tips in these systems. We are confident that the FPTA constitutes a unique technology for spee\-ding up the artificialization and digitalization of topological quantum matter.

\section*{Acknowledgements}

This work was supported by the National Natural Science Foundation of China (NSFC) (11904399) and the Open Research Fund from State Key Laboratory of High Performance Computing of China (HPCL) (Grant No. 201901-09).

G.-Y.~H. and B.~L. contributed equally to this work.

\appendix 

\section{Further Details for Case-Study D}\label{sec:AppendixB}

The nanowire electrons are coupled to the 2DEG under the MJ by means of the Hamiltonian: 
\bea
{\cal H}_{\rm MJ\mbox{-}TSC} =\sum_{\sigma=\uparrow,\downarrow}\sum_{s=1,\ldots,4}\sum_{n,\bar{m}}\big(c_{n\sigma;s}^{\dag}T_{n\bar{m}}^sc_{\bar{m}\sigma}+{\rm H.c.}\big),\qquad
\label{eq:MJ-TSC Hamiltonian}
\eea

\noi with $s=1,2,3,4$ labelling the corresponding TSC. For the remainder, we assume that the matrix elements $T_{n\bar{m}}^s\in\mathbb{R}$ do not depend on the nanowire index, and are generally nonzero only for nearest and next-nearest neighbours, with strengths $T_1$ and $T_2$, respectively. 

We assume that the potential energies $U_{\bar{n}}$ are positive/negative and sufficiently large to guarantee that the occupations of the respective sites are practically pinned to zero/one. Therefore, the 2DEG sites underneath the MJ function as steppingstones for the nanowire electrons to hop from one nanowire to another by only virtually oc\-cu\-pying the junction's sites. Hence, in the low-energy sector, i.e., with $\varepsilon\ll U_{\bar{n}}$ $\forall\bar{n}$, we find the inter-nanowire Hamiltonian:
\bea
{\cal H}_{\rm TSC}=\sum_{\sigma=\uparrow,\downarrow}\sum_{s\neq s'}^{1,\ldots,4}\sum_{n,m}^{\rm TSC}c_{n\sigma;s}^{\dag}{\rm T}_{nm}^{ss'}c_{m\sigma;s'},\quad
\eea

\noi with $n$ and $m$ being restricted to the sites comprising the TSC nanowires.

To determine the inter-nanowire tunnel couplings, we integrate out the electronic degrees of freedom of the MJ by means of the Dyson self-energy formula, and expand the resul\-ting inter-nanowire tunnel couplings up to se\-cond order with respect to $t_1$ and $t_2$. These steps yield:
\bea
{\rm T}_{nm}^{ss'}&\approx&
\sum_{\bar{n},\bar{m}}T_{n\bar{n}}^s
\bigg(-\frac{\delta_{\bar{n}\bar{m}}}{U_{\bar{n}}}+\frac{t_{\bar{n}\bar{m}}}{U_{\bar{n}}U_{\bar{m}}}\no\\
&&-\sum_{\bar{r}}\frac{t_{\bar{n}\bar{r}}t_{\bar{r}\bar{m}}}{U_{\bar{n}}U_{\bar{r}}U_{\bar{m}}}\bigg)
e^{\pi i\int_{\bm{R}_{\bar{n}}}^{\bm{R}_{\bar{m}}}\bm{A}\cdot{\rm d}\bm{r}/\Phi_0}T_{\bar{m}m}^{s'}\label{eq:Tcouplings}
\eea

\noi where the site indices $\bar{n},\bar{m}$ and $\bar{r}$ are restricted to the sites underneath the MJ. We note that $s\neq s'$ in the above, since we consider that the intra-nanowire processes do not contribute, because their effects can be absorbed by suitably redefining the chemical potentials of the nanowires. In order to obtain Eq.~\eqref{eq:Tcouplings}, we first performed a gauge transformation in order to eliminate the Peierls factors from Eq.~\eqref{eq:MJ Hamiltonian} and transfer them to Eq.~\eqref{eq:MJ-TSC Hamiltonian}. 

In this low-energy regime, we further appro\-xi\-ma\-te eve\-ry nanowire electronic operator as $c_{n\sigma;s}=u_{n\sigma;s}\gamma_s$, with $u_{n\sigma;s}$ denoting the electronic components of the state vector of the $s$-th MZM. The Majorana operators adhere to the relation $\{\gamma_s,\gamma_{s'}\}=\delta_{ss'}$, which leads to the nor\-ma\-li\-za\-tion condition $\sum_{\sigma,n}|u_{n\sigma;s}|^2=1/2$ $\forall s=1,2,3,4$. Assuming that the nanowires are cha\-rac\-te\-ri\-zed by ge\-ne\-ral\-ly-different superconducting phases $\phi_s$, but are otherwise identical, brings us to the expression $u_{n\sigma;s}=e^{-i\phi_s/2}/\sqrt{2}$ for both $\sigma=\uparrow,\downarrow$. This wavefunction structure is obtained for a magnetization lying in the $xz$ plane and a SOC orien\-ted in the $y$ axis, which together further imply the pre\-sen\-ce of chiral symmetry for each nanowire with an operator $\Pi_s=\tau_xe^{i\phi_s\tau_z}$, in the spinor basis $c_{n;s}^{\dag}=(c_{n\uparrow;s}^{\dag},\,c_{n\downarrow;s}^{\dag},\,c_{n\uparrow;s},\,c_{n\downarrow;s})$. The state vectors correspon\-ding to the $u_{n\sigma;s}$ above are eigenstates of $\Pi_s$, and possess the same chi\-ra\-li\-ty $\Pi=+1$ for $\phi_{1,2,3,4}=0$. This convenient choice preserves the ge\-ne\-ra\-li\-ty of our results~\cite{Meyer,Sakurai}.

Concluding, we briefly comment on the limiting case $\Phi_{\rm MJ}=0$, which generally leads to a nonzero $I_{\rm Grid}$. The key difference compared to the limiting case analyzed in the main text, is that in this case the $g_{1;1}$ component of the $\bm{g}_1(\phi_{21},\phi_{43})$ vector becomes independent of the synthetic momenta $\phi_{21,43}$. Hence, a Dirac mass is straightforward to define and the resulting analysis of the Weyl points boils down to inferring the conditions that render the mass zero as a function of the various parameters. Finally, the ABS energy spectrum retains its $4\pi$-periodicity with respect to each $\phi_{21,43}$ phase difference.

\end{document}